\documentclass[a4paper]{jpconf}
\usepackage{lmodern}        
\usepackage[T1]{fontenc}
\usepackage[utf8]{inputenc} 
\usepackage{graphicx}  
\usepackage[pdftex,bookmarksopen,bookmarksnumbered,linktoc=all,hidelinks]{hyperref}

\begin{document}
\title{Timing analysis of rotating radio transients discovered with MeerKAT}

\author{Thulo Letsele$^{1}$, Mechiel Christiaan Bezuidenhout$^{2,1}$, and the MeerTRAP collaboration}

\address{$^1$ Centre for Space Research, North-West University, Potchefstroom 2531, South Africa}
\address{$^2$ Department of Mathematical Sciences, University of South Africa, Cnr Christiaan de Wet Rd and Pioneer Avenue, Florida Park, 1709, Roodepoort, South Africa}

\ead{thuloletsele1999@gmail.com}
\begin{abstract}
Pulsars are rapidly rotating neutron stars that emit pulses of radiation at regular intervals, typically ranging from milliseconds to seconds. The precise recording and modelling of the arrival times of pulsar emission is known as timing analysis. Rotating radio transients (RRATs) are a subclass of pulsars that emit pulses very sporadically. Because of the sparse pulse times of arrival (ToAs) typically available for these sources, they are much more difficult to time than regular pulsars, to the extent that few RRATs currently have coherent timing solutions. In this work, we present the results of timing analyses for four RRATs discovered by the MeerTRAP transient survey using MeerKAT. We incorporated additional pulse ToAs from each source that have been detected since their original analysis. We confirmed the known timing solution for PSR~J1843$-$0757, with a period of $P=2.03$ seconds, and a period derivative of $\dot{P}=4,13\times10^{-15}$. However, our analysis did not comport with the solution of MTP0005, which we conclude may have been mistakenly identified with the known PSR~J1840$-$0815 in the original analysis. Finally, the spin period for MTP0007 was determined to be $1.023(1)$ seconds using a brute-force period fitting approach.
\end{abstract} 
\section{Introduction}
The death of a massive star ($>$8~M$_{\odot}$) as a supernova is thought to leave behind a compact remnant, often in the form of a rotating and magnetised neutron star known as a pulsar. In addition to regular pulsars, Rotating Radio Transients (RRATs) are a subpopulation of pulsars that display significant variation in their radio emission, and often switch off for many pulse periods. The behaviour of switching off for seconds or even a year at some time is known as nulling. Many RRATS are visible for a short time in between nulls, which makes them very difficult to detect. Due to RRATs' intermittency, they are easier to detect using single pulse searches than periodicity searches \cite{1}.\\
\\
Single pulse searching is a method used to detect and analyse individual pulses or bursts of radio emission. It is used for sources that do not exhibit regular pulsations by identifying significant deviations from the background noise level in the time-domain signal. It requires specialised algorithms and hardware to handle the large amounts of data. An example of such an algorithm is $\textsc{AstroAccelerate}$ \cite{2}, used by the MeerTRAP (more TRANsients and Pulsars) commensal transient survey using the MeerKAT radio telescope \cite{26}.\\
\\
The precise measurement and analysis of the arrival times of pulses emitted by pulsars is known as pulsar timing. There are several factors that need to be accounted for in a so-called timing model. These include intrinsic properties of the pulsar, such as its period, period derivative, and position in the sky. Additionally, it must also consider the effects of the pulsar's motion, its orbit around a companion star, the Earth's motion, and the interstellar medium through which the pulsar signals propagate \cite{3}. Other factors include relativistic effects caused by the curvature of spacetime near massive objects and the gravitational interaction with other bodies in the system. RRATs can be timed using the same methods as for regular pulsars if a sufficient number of single pulses can be detected.\\
\\
The first twelve Galactic fast transients discovered by MeerTRAP were presented in \cite{13}. They derived a full timing solution for one source, MTP0001 (J1843$-$0757), and calculated approximate spin periods for another three sources using an arrival time differencing method\footnote{\url{https://github.com/evanocathain/Useful\_RRAT\_stuff}}. One source, MTP0005, was initially identified as a newly discovered transient, but was later concluded to be from PSR~J1840$-$0840, a pulsar discovered in the Parkes Multibeam Pulsar Survey \cite{27}.\\
\\
For this paper, we performed timing analysis of four transient sources presented in \cite{13}: MTP0001, MTP0005, MTP0007 and MTP0011. Since that publication, MeerTRAP has detected more pulses from these four sources, meaning ToAs will be added to the data set. Using these additional ToAs, we attempted to reproduce the timing solution for MTP0001 and MTP0005 (PSR~J1840$-$0840), and to derive a new timing solution for the other two sources. 
\section{Data processing}
\begin{figure}[!ht]
\centering
\includegraphics[width=.999\textwidth]{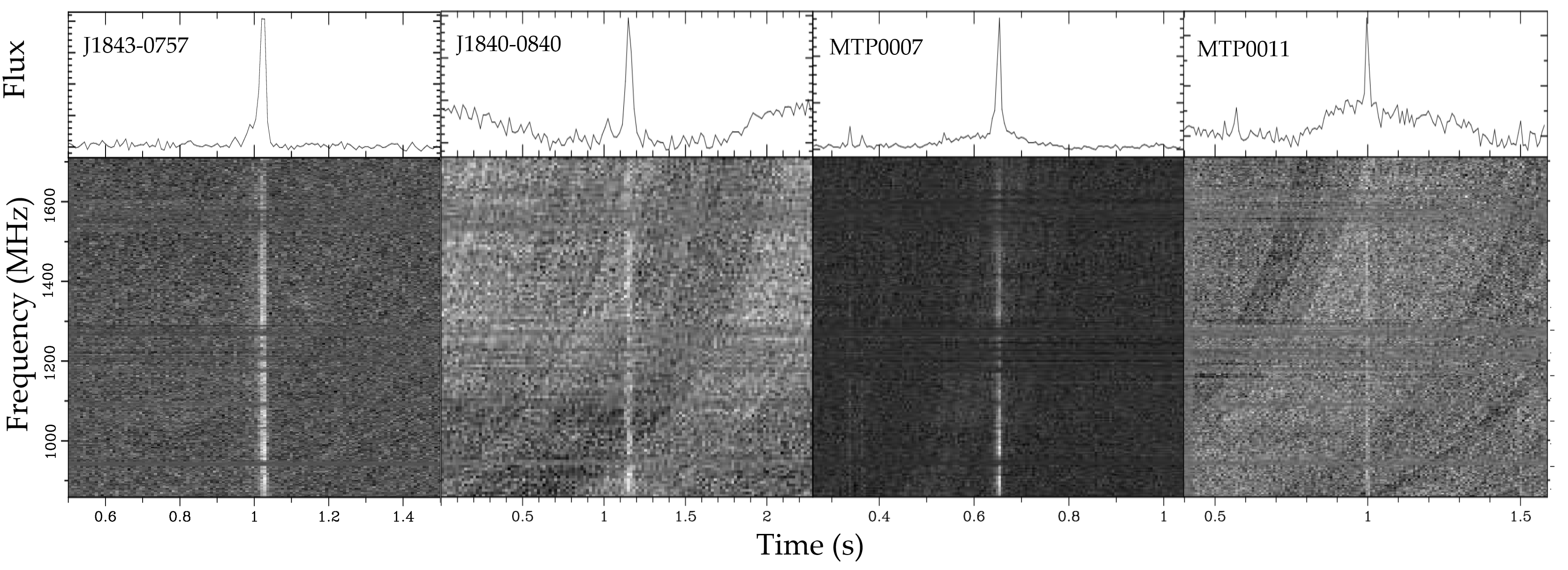}
\caption{(\emph{Top}) \label{fig2}Average profile and (\emph{Bottom}) dynamic spectrum for each RRAT. The sweeping features in the dynamic spectra are zero-DM RFI signatures.}
\end{figure}

\begin{figure}[!ht]
\centering
\includegraphics[width=0.49\textwidth]{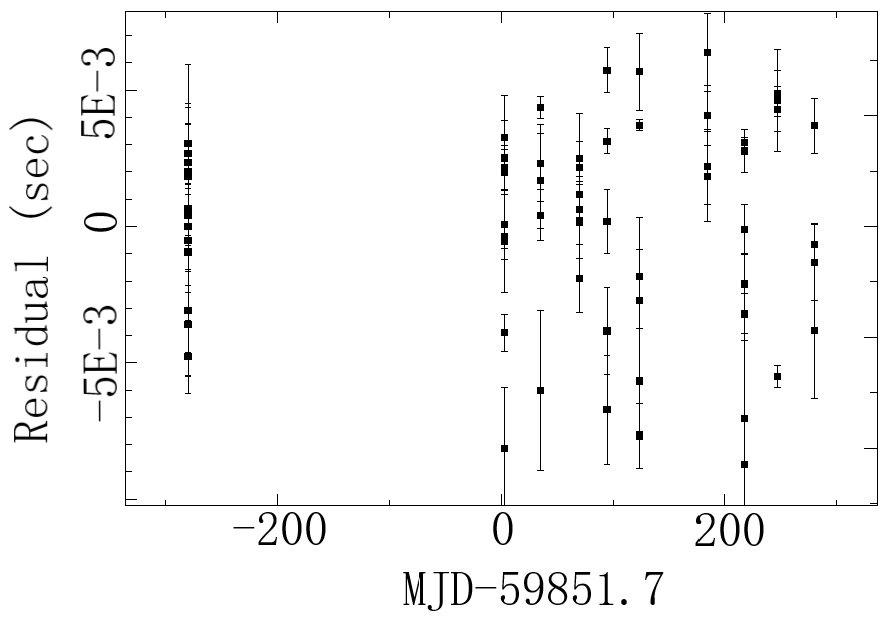}
\includegraphics[width=0.49\textwidth]{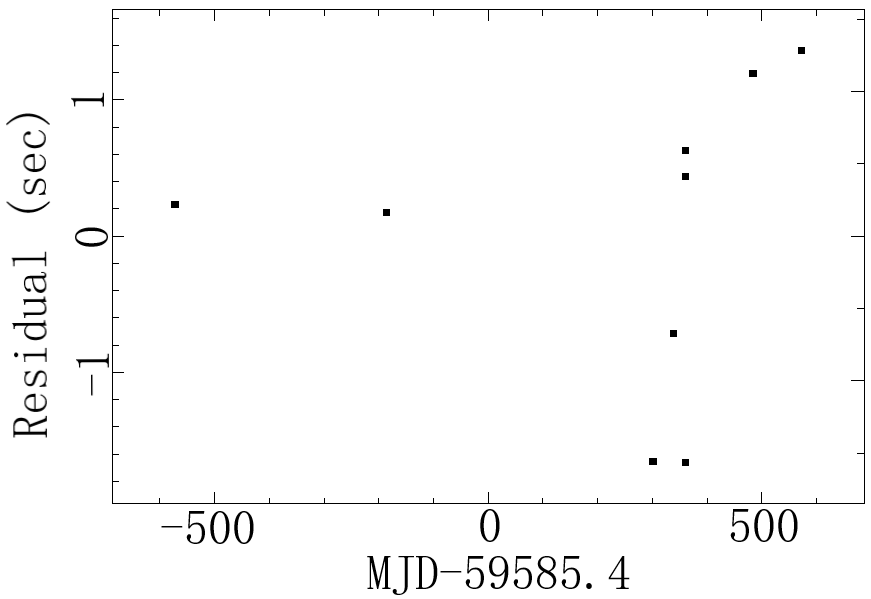}
\caption{\label{fig:15} (\emph{Left}) ToA residuals after fitting for the timing model for PSR~J1843$-$0757. (\emph{Right}) ToA residuals after fitting for the timing model for PSR~J1840$-$0840. The error bars on the PSR~J1840$-$0840 residuals are too small to show on this scale.}
\end{figure}

MeerTRAP uses the channelised time series data from the Filterbank BeamFormer User Supplied Equipment (FBFUSE) beamformer \cite{20}, and processes them in real time on the Transient User Supplied Equipment (TUSE) cluster. Candidate pulses identified using $\textsc{AstroAccelerate}$ are stored as filterbank files, with each file containing the data for one pulse. We identified new pulses from each source by matching candidates to the known source coordinates and dispersion measure (DM), which is the integrated electron column density from the telescope to the source. The first step of the analysis involved de-dispersing each single-pulse filterbank file using $\textsc{DSPSR}$\footnote{\url{http://dspsr.sourceforge.net/}}. The result of this process was a corresponding $\textsc{PSRCHIVE}$ format file for each filterbank, containing the de-dispersed data.\\ 
\\
The pulse typically contains narrow-band radio frequency interference (RFI). In order to clean the pulse archive, the $\texttt{clfd}$\footnote{\url{https://github.com/v-morello/clfd}} automatic RFI excision algorithm was used \cite{28}. After that, the dynamic spectrum of each file was manually reviewed to identify any positive detections. If a pulse was detected in multiple beams, the one with the highest signal-to-noise ratio ($S/N$) was chosen. The number of positive detections for each source is listed in Table~\ref{tab:detections}. All positive detections were combined to increase the S/N (see Figure \ref{fig2}). A standard profile was then created using the \texttt{PSRCHIVE} tool \texttt{paas}, and ToAs were extracted using $\texttt{pat}$. $\textsc{TEMPO2}$\footnote{\url{http://www.atnf.csiro.au/research/pulsar/tempo2}} \cite{25} was then used to  fit the timing model. Figure \ref{fig:15} shows the residuals between the timing model and the observed ToAs for MTP0001 and MTP0005, respectively.

For the sources without an existing timing solution, we used the \textsc{RRATsolve}\footnote{\url{https://github.com/v morello/rratsolve/tree/master}} brute-force algorithm, which determines the largest possible period that can divide all the time intervals provided between the first and the subsequent input ToAs. The algorithm chooses a geometrically spaced multiplicative factor between the consecutive trials period, so that one point falls within the expected 1-sigma confidence interval for the true period. The algorithm then selects points that are close enough to the true period and provides an analytical solution.

\begin{center}
\begin{table}[h]
\caption{\label{tab:detections}Number of detections and S/N of the average pulse for each source, as well as the derived timing parameters where available. The values indicated in brackets correspond to 1-$\sigma$ uncertainty estimates on the derived best-fit timing model.}
\centering
\begin{tabular}{@{}*{6}{lrrrll}}
\br
Source & DM (pc~cm$^{-3}$) & No. of detections & S/N & P (s) & $\dot{P}$ (s/s)\\
\mr
J1843$-$0757 & 254 & 46 & 82 & 2.0319400853(2) & 4.134(8)$\times$10$^{-15}$\\
J1840$-$0840 & 299 & 9 & 36 & 1.096441140382(4) & 2.4264(1)$\times$10$^{-15}$\\
MTP0007 & 33 & 22 & 62 & 1.023(1) & \\
MTP0011 & 174 & 7 & 21 & & \\
\br
\end{tabular}
\end{table}
\end{center}

\section{Results}
A good timing solution with satisfying residuals ($\mathrm{rms}=3.7~\mathrm{ms}$) for PSR~J1843$-$0757 was obtained. The period of $P$ is measured to be $2.0319400853(2)$ seconds, and the period derivative of $\dot{P}$ is $4.134(8) \times 10^{-15}$. Timing parameters derived using TEMPO2, with 1-$\sigma$ errors indicated in brackets, have been demonstrated to be accurate down to sub-ns level \cite{2006edwards}. This solution aligns with that reported in \cite{13}. However, for MTP0007 and MTP0011, there were too few viable ToAs to obtain a timing solution.\\
\\
For PSR~J1840$-$0840, the residuals were very large (see Figure \ref{fig:15}) and we could not get a good fit to the data, indicating that there was a mismatch between the MeerTRAP detections observations and the known timing solution, as reflected on the ATNF pulsar catalogue\footnote{\url{https://www.atnf.csiro.au/research/pulsar/psrcat/}}, which is listed in Table~\ref{tab:detections}. We also tried to derive an independent pulse period using \textsc{RRATsolve}, but the algorithm did not converge on an estimate. This result indicates that the identification of MTP0005 with PSR~J1840$-$0840 in \cite{13} may have been mistaken, and that there could be two sources with similar DM nearby one another. We note that, of the nine detections included in our sample, four were detected in the MeerKAT incoherent beam, and only five in tied-array beams. The tied-array beam detections were also spatially distant from one another (by up to 21 arcminutes), and from the recorded position of PSR~J1840-0815 (by up to 13 arcminutes). It is possible that these represent side-lobe detections of the same source, but---in light of the timing analysis---we consider it more likely that two or more distinct sources are present. More observations are required to establish this for certain, particularly in order to localise the source more precisely, either using difference imaging or a beamformed localisation algorithm \cite{29}.\\
\section{Conclusion}
Since neither MTP0007 nor MTP0011 are known pulsars, existing ephemerides cannot be used to time them. Therefore, brute-force methods must be used to obtain period measurements for them. Based on the measurements obtained through \textsc{RRATsolve}, the period for MTP0007 was calculated to be $1.023$$\pm$$0.001$ seconds. In addition, \textsc{RRaTsolve} was used for MTP0011, but the ToAs were too sparsely distributed for it to work. It is possible that these sources are infrequently active, resulting in the insufficient number of pulses for analysis. Additionally, it is worth noting that MeerKAT may spend relatively little time directly pointing towards the source coordinates, potentially leading to reduced sensitivity in the primary beam. Further investigations and observations are required to explore these sources in more detail and potentially capture more pulses for future timing solutions.\\

\section*{Acknowledgements}
The MeerTRAP collaboration acknowledges funding from the European Research Council (ERC) under the European Union’s Horizon 2020 research and innovation programme (grant agreement No 694745).
\section*{References}
\bibliographystyle{iopart-num}
\bibliography{iopart-num}

\end{document}